\documentclass[conference]{IEEEtran}
\usepackage{graphicx}
\usepackage{textgreek}
\usepackage{float}
\usepackage{amsmath}
\usepackage{cite}
\IEEEoverridecommandlockouts % Required for inserting images

\begin{document}

\title{FACE2FEEL: EMOTION-AWARE ADAPTIVE USER INTERFACE}
\author{
\IEEEauthorblockN{Alihan Hadimlioglu, Ph.D.}
\IEEEauthorblockA{
Department of Computer Science\\
Texas A\&M University-Corpus Christi \\
TX, USA\\
Email: alihan.hadimlioglu@tamucc.edu}
\and
\IEEEauthorblockN{Siddharth Linga}
\IEEEauthorblockA{
Department of Computer Science\\
Texas A\&M University-Corpus Christi \\
TX, USA \\
Email: siddharthlinga2808@gmail.com}
}

\maketitle
\begin{abstract}
This paper presents Face2Feel, a novel user interface (UI) model that dynamically adapts to user emotions and preferences captured through computer vision. This adaptive UI framework addresses the limitations of traditional static interfaces by integrating digital image processing, face recognition, and emotion detection techniques. Face2Feel analyzes user expressions utilizing a webcam or pre-installed camera as the primary data source to personalize the UI in real-time. Although dynamically changing user interfaces based on emotional states are not yet widely implemented, their advantages and the demand for such systems are evident. This research contributes to the development of emotion-aware applications, particularly in recommendation systems and feedback mechanisms. A case study, "Shresta: Emotion-Based Book Recommendation System," demonstrates the practical implementation of this framework, the technologies employed, and the system's usefulness. Furthermore, a user survey conducted after presenting the working model reveals a strong demand for such adaptive interfaces, emphasizing the importance of user satisfaction and comfort in human-computer interaction. The results showed that nearly 85.7\% of the users found these systems to be very engaging and user-friendly. This study underscores the potential for emotion-driven UI adaptation to improve user experiences across various applications.
\end{abstract}

\begin {IEEEkeywords}
Adaptive user interface, Component-based adaptive UI, Digital image processing, Frames, Deep Face, Emotion detection, Recommendation engine, Human-computer interaction.
\end{IEEEkeywords}

\section{Introduction}
In this era of technological advancement, where individuals rely heavily on electronic gadgets to accomplish daily tasks, significant time is spent engaging with mobile applications and websites. Studies conducted by European Psychiatry \cite{suicide} and the Journal of Health and Social Behavior \cite{Adolesent-health} indicate that a dominant causes of teenage suicides are social isolation and excessive screen time. Therefore, emotion-aware adaptive UIs offer an effective solution to address these issues.

Social interaction plays a crucial role in mitigating negative emotion such as sadness and depression. In such circumstances, individuals can rely on their social support, including reciving advice, sharing laughter, and engaging in humor. On the other hand social isolation deprives individuals of these critical interactions, preventing them to cope with emotional distress. Suppose this person uses a social media application like Instagram. If Instagram integrates an emotion-aware adaptive UI with features such as emotion detection, dynamic UI, and emotion tracking, it could identify the user’s emotional state and respond accordingly. Through such an implementation, instead of displaying content related to accidents or conflicts, the system could prioritize humorous, inspirational, or uplifting videos. This type of UI has the potential to improve the mood of its users. Furthermore, if AI is integrated into this design, the system could analyze emotional patterns over time. For instance, if the user frequently exhibits anger, the system could recommend anger management resources or tailored content to address this behavior.

Beyond social media and entertainment, emotion-aware adaptive UIs can play a vital role in professional growth and learning platforms. Consider a student facing several impending assignment deadlines. If the learning platform detects that the student is feeling sad, it could suggest completing simpler tasks first to improve their mood before tackling more challenging assignments. Conversely, if the system detects a neutral or happy emotional state, it could recommend addressing complex tasks initially. Similarly, for working professionals, such a system could prioritize tasks based on their emotional state, ensuring optimal productivity and mental well-being.

Adaptive UIs are not limited to recommendation systems; they also enhance the user interface's visual appeal. For example, if the user is angry, the background color could change to bright or soothing tones. Additionally, a calming sound might play in the background, and subtle animations could be introduced to help the user feel better. Such features demonstrate how adaptive UIs can significantly transform the current technological landscape.

Face2Feel is an emotion-aware adaptive UI interface designed to dynamically adapt to users' emotions. This system emphasizes creating a user-friendly and visually appealing environment tailored to the user's needs and preferences. By leveraging computer vision concepts and Digital Image Processing (DIP), the system performs emotion detection on videos captured through webcams or phone cameras. Based on the detected emotion, the UI reacts and dynamically adjusts itself to enhance the user experience \cite{EBAUI}.

Face2Feel extends beyond the concept of adaptive UIs with default implementations by incorporating customization features that allow users to tailor the interface according to their preferences. This ensures that the system aligns with the user's specific needs, rather than being limited to changes dictated by developer-defined interests. Without this customization capability, the core principle of an emotion-based adaptive UI would remain unjustified, as the system would fail to provide a truly personalized experience.

Furthermore, the system integrates an emotion tracking mechanism that monitors the user's emotions over a period of time. This feature enables the application to understand the emotional patterns and states the user has experienced recently, allowing it to adapt dynamically and provide a more contextualized response. The incorporation of AI-driven enhancements further elevates the system's functionality, enabling it to make intelligent adjustments and deliver a superior user experience\cite{intelligent_UI_design}.

\section{Literature Survey}
\subsection{Emoticontrol: Emotions-based Control of User-Interfaces Adaptations}
The document presents Emoticontrol, a system that modifies user interfaces (UIs) in accordance with emotions utilizing Model-Free Reinforcement Learning (MFRL). It emphasizes the significance of taking human emotions into account in UI design to enhance Quality of Experience (QoE) and prevent discomfort or system malfunctions. The method proactively modifies UIs by identifying emotions through facial analysis and employing MFRL to fine-tune adaptations, guaranteeing effective task performance and user contentment\cite{Emoticontrol}.

Deployed in a mobile application for emergency evacuation training, the system directs users toward safety while regulating their emotional conditions. Technologies such as facial recognition, reinforcement learning, and adaptive UI structures drive the solution. Experiments confirm its efficacy, surpassing conventional rule-based techniques and balancing user satisfaction with system efficiency. This work illustrates the potential of AI-enhanced adaptive UIs in emotionally intense situations.

Limitations \cite{Emoticontrol}:

\begin{itemize}
    \item Dependency on MFRL: Although MFRL learns dynamically, its effectiveness is significantly dependent on the Q-table initialization, which might restrict adaptability in situations not foreseen during training.
    
    \item Scalability Challenges: The system manages emotions in real-time, which may not scale effectively for larger or distributed user populations without considerable computational resources.
    
    \item Limited Emotion Categories: The emphasis on Ekman’s six fundamental emotions omits subtle emotional states that could influence the efficiency of UI adaptation.
    
    \item Experiment Scope: The application is designed specifically for evacuation training, restricting its applicability to wider areas of UI adaptation.
    
    \item Ethical and Privacy Concerns: Even with anonymized data management, real-time tracking of emotions could provoke concerns regarding user consent and potential data misuse in real-world applications.
\end{itemize}
\subsection{Model-based adaptive user interface based on context and user
experience evaluation}\cite{Model-based-UI}
The document presents a model-based adaptive UI approach, executed via the A-UI/UX-A tool in the Mining Minds platform. The system utilizes technologies like the Laravel PHP framework, the Protégé editor for ontology models, and the Semantic Web Rule Language (SWRL) for generating inference rules. Contextual inputs are obtained from various multimodal data sources, including sensors and feedback systems, and are processed by reasoning engines to create UIs dynamically. Adaptive modifications are applied according to user cognition, device capabilities, and environmental aspects, ensuring personalized user experiences\cite{Model-based-UI}.

Limitations:

\begin{itemize}
    \item Complexity in Rule Creation: The requirement for expert-level rule creation in the methodology being discussed can become a hindrance.
    \item Aesthetic Limitations: User interfaces generated automatically do not possess the visual appeal of those crafted by designers. 
    \item Frequent Adaptations: Repeated changes to the user interface can interfere with user learning. 
\end{itemize}

\section{Implementation}
For an adaptive user interface, there are 2 main aspects to implement, This section elaborates on these aspects.
\subsection{User Interface}
The User Interface (UI) serves as the primary interactive interface between the computer and the user. The concept of Adaptive User Interfaces (AUIs) is predominantly focused on this section, as it is the only part of the application with which the user directly interacts. There are several approaches to making UIs adaptive and interactive, among which Component-Based Adaptive UIs (CBAUIs) are considered highly effective and efficient for implementing dynamically changing UIs. By dividing the page into components, necessary components can be modified individually as required, simplifying the development process and reducing the need for complex methods. Additionally, implementing customization features becomes significantly easier within such frameworks \cite{component-based-AUI,CBAUI-best-approach}.

Using a CBAUI approach enhances scalability, reusability, modularity, flexibility, and consistency. There are several languages and frameworks available to facilitate these tasks, including HTML/CSS, React, Angular, Preact, and others. The selection of a framework depends on the intensity of the project, dataset requirements, complexity, and the limitations of the chosen framework.

For instance:
\begin{itemize}
    \item If HTML/CSS alone are used, it becomes challenging to implement an AUI due to limited capabilities in making components dynamic. In such cases, additional frameworks, such as JavaScript, are required to enable dynamic adaptability in the application\cite{webcomponents}\cite{dynamic-html}.
    \item For applications with manageable data and moderate AUI requirements, ReactJS can be a suitable choice, offering a component-based structure and efficient state management.
    \item For applications with complex datasets and higher demands for scalability and flexibility, Angular may be a more appropriate choice, providing a robust framework with extensive built-in features to handle such complexities.
\end{itemize}
Changes in the User Interface (UI) are dynamically made based on the user's emotions. For instance, in a default setting, when a user is angry, they might prefer content that is calm, soothing, or humorous. Consider a scenario where a user, feeling frustrated and angry, opens Instagram and begins scrolling through reels. If Instagram employs an emotion-aware Adaptive UI (AUI), it could dynamically adjust the feed to display more calming or humorous content, helping to alleviate the user's mood\cite{Mahdi}.

Additionally, if the system incorporates emotion tracking features, it could analyze the user's emotional patterns over time. For example, if the system observes that the user frequently experiences anger, it could proactively recommend anger management resources or related content in the Instagram feed. This approach not only enhances the adaptability of the UI but also helps the user regulate their mood and feel better, creating a more personalized and beneficial experience\cite{emotion-tracking}.

\subsection{Emotion detection}
\begin{figure}[H]
\centering
\includegraphics[width=8cm]{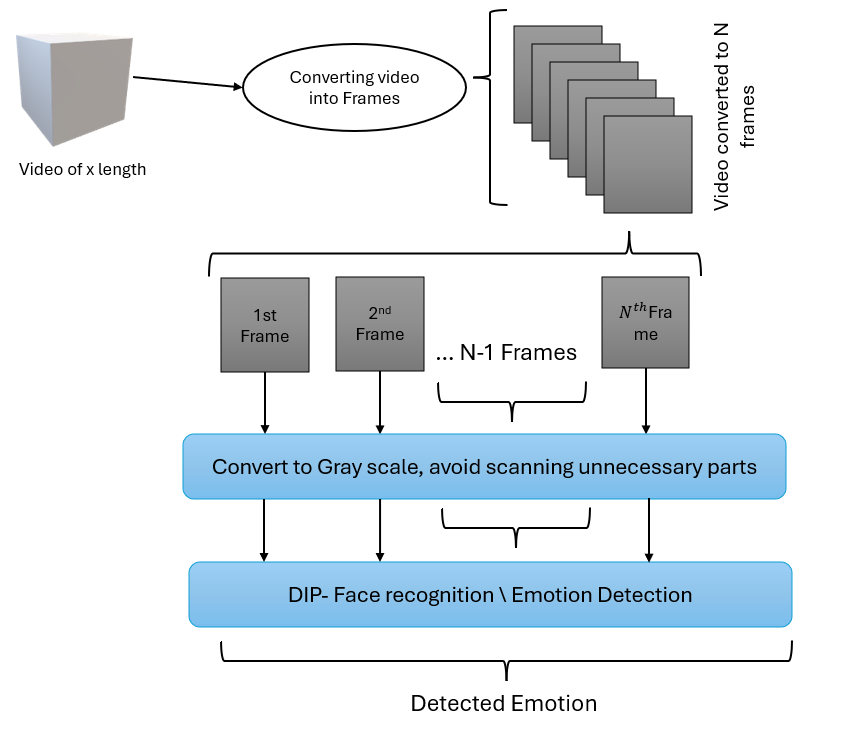}
\caption{working of DIP and emotion detection}
\label{fig:f8}
\end{figure}

For emotion-aware AUIs, the primary data source is video input, typically acquired through inbuilt cameras on devices like smartphones and tablets or webcams on laptops and computers. However, performing video processing directly for emotion analysis or face recognition is computationally intensive and complex. A practical solution to this challenge involves converting video into frames. As illustrated in Figure \ref{fig:f8}, the video input is segmented into individual frames—essentially images captured at fixed time intervals. Digital Image Processing is then applied to these frames.

Frames are discrete snapshots of a video taken at regular intervals. For example, for a 5-second video with a snapshot captured every 0.01 seconds, the system generates 500 frames. DIP, combined with DNNs—especially Convolutional Neural Networks (CNNs)—is utilized to perform emotion detection and face recognition on each frame. This approach ensures accurate results and enables the seamless functioning of the system. As shown in Figure \ref{fig:f8}, the video is converted into frames, and each frame is analyzed to detect user emotions and improve the adaptability of the UI\cite{ED-DWT,ED-opencv,ED-transferlearning-DL,ED-deepface}.

For emotion detection we can use Digital Image Processing (DIP), face recognition and emotion detection are critical features that play a significant role in implementing Adaptive User Interfaces (AUIs). By leveraging concepts such as Deep Neural Networks (DNNs), tasks like face and emotion detection can be performed effectively. As discussed in the case study, numerous libraries and frameworks are available to help achieve the desired results.
\subsection{Optimization and time complexity}
As illustrated in Figure \ref{fig:f8}, the video input is divided into individual frames, and Digital Image Processing (DIP) is performed on each frame to determine the user's emotion. However, implementing this process traditionally without optimization measures results in a suboptimal application with the worst-case time complexity. To address this challenge, the following optimization measures are proposed:

\begin{itemize}
    \item \textbf{Grayscale Conversion:}
    
    Each frame is converted to a grayscale image, reducing computational complexity. A grayscale image has intensity values represented by a single channel, compared to RGB images, which have three channels (Red, Green, and Blue). This conversion simplifies tasks such as face and emotion detection, as well as feature extraction. Specifically, processing grayscale images requires approximately one-third of the computational effort compared to RGB images.

    \item \textbf{Multi-Processing and Section Selection:}
    
    Without multi-processing, the application experiences increased computational load and worst-case time complexity, making it non-optimal. By implementing multi-processing, the time complexity is significantly reduced, as tasks are distributed across multiple processors, enabling faster execution. Additionally, scanning the entire image for face and emotion detection is unnecessary. By focusing only on relevant sections of the image (e.g., areas containing a face) and skipping irrelevant parts, the application achieves further optimization\cite{multiprocessing-timecomplexity}\cite{multiprogramming}.
\end{itemize}

To simplify calculations, the following parameters are defined:
\begin{itemize}
    \item $X$: Length of the video input in seconds.
    \item $N$: Number of frames generated from the video (e.g., if a frame is captured every 0.01 seconds, $N = \frac{X}{0.01}$).
    \item $M$: Number of pixels in an image or the frame size.
    \item $P$: Number of processors available for performing the task.
    \item $\alpha$: Proportion of pixels analyzed (\(\alpha M\)), where \(\alpha < 1\), to exclude irrelevant sections.
\end{itemize}

\subsubsection{When Multiprocessing and Optimization Are Applied}
\begin{itemize}
    \item \textbf{Grayscale Conversion:}

    Grayscale conversion involves iterating through $\alpha M$ pixels per frame, resulting in a time complexity of $O(\alpha M)$.

    \item \textbf{Face Detection:}

    After grayscale conversion, $\alpha M$ pixels are analyzed for face and emotion detection, yielding a time complexity of $O(\alpha M)$.

    \item \textbf{Multiprocessing:}

    With $P$ processors, $N$ frames are distributed equally. The time complexity per processor is:
    \[
    T_{\text{multiprocessing}} = \frac{N}{P} \cdot O(\alpha M)
    \]
    The effective total time complexity of the application is:
    \[
    T_{\text{total-multiprocessing}} = O\left(\frac{N \cdot \alpha M}{P}\right)
    \]
\end{itemize}

\subsubsection{Without Multiprocessing}
\begin{itemize}
    \item \textbf{Grayscale Conversion:}

    The entire image containing $M$ pixels is processed, resulting in a time complexity of $O(M)$.

    \item \textbf{Face Detection:}

    Similarly, all $M$ pixels are analyzed for face detection, yielding a time complexity of $O(M)$.

    \item \textbf{Sequential Processing:}

    For $N$ frames, the total time complexity becomes:
    \[
    T_{\text{total-sequential}} = N \cdot O(M)
    \]
\end{itemize}

\subsubsection{Comparison and Conclusion}

By comparing both time complexities, it is evident that employing optimizations such as multi-processing and section selection significantly reduces the computational load and enhances the efficiency of the emotion detection process. These measures ensure the feasibility of generating Emotion-Based Adaptive UIs in a highly efficient and scalable manner.

\section{Case Study}

\subsection{Introducing "Shresta: Emotion-based Book recommendation system" }
\begin{figure}[H]
\centering
\includegraphics[width=7cm]{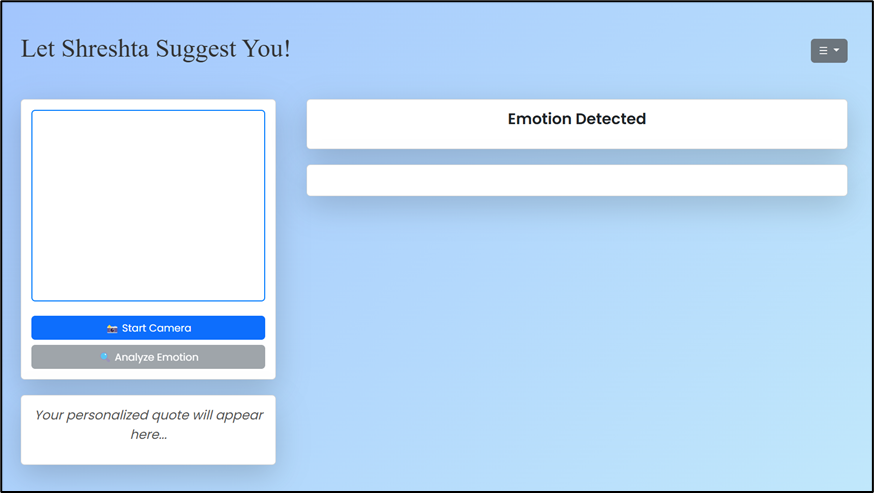}
\caption{Glimpse of the system}
\label{fig1}
\end{figure}

Using the concept of Adaptive UI, this engine recommends books to users based on their emotions while also offering a dynamic UI that adapts to the detected emotion. The name “Shreshta” derived from Sanskrit, means “the best,” symbolizing an exceptional system where emotion and literature converge. This engine not only recommends books but also dynamically changes the background based on user emotions, provides animations, and displays emotion-based quotes. Since books are deeply tied to emotions, this system offers a real-time application that combines emotional adaptation with meaningful recommendations. Figure \ref{fig1} shows the initial view of the application.  

\subsection{System Architecture and Implementation}
This project is built upon three core aspects:
\begin{itemize}
    \item A dynamic front-end for user interaction and book recommendations.
    \item Emotion detection and server-side processing to analyze user emotions.
    \item Book recommendations dynamically adjusted based on detected emotions.
\end{itemize}
The overall system architecture, including the frontend, JavaScript component and the Python backend, is illustrated in Figure 3.
\subsubsection{Frontend}
The frontend of the system is developed using HTML, CSS, and Bootstrap. Key technologies and design elements include:\cite{dynamic-html,webcomponents}
\begin{itemize}
    \item \textit{HTML Components}: Utilized containers, grids, and cards to align text and arrange various elements effectively.
    \item \textit{CSS Styling}: Added vibrant colors, animations (e.g., emoji rain), and attractive layouts to enhance user experience.
    \item \textit{Bootstrap Integration}: Incorporated features such as a hamburger menu, dashboards, customization sections, and buttons for a more interactive and visually appealing application\cite{bootstrap}.
    \item \textit{Recommendation Engine}: Designed to display book recommendations in a scrollable format, showcasing multiple options\cite{Mahdi}.
\end{itemize}

\subsubsection{JavaScript for Interfacing}
JavaScript acts as the bridge between the front-end HTML/CSS pages and the Python server. It performs the following tasks:\cite{dynamic-html,webcomponents}
\begin{itemize}
    \item \textit{Communication}: Sends user actions and input data from the frontend to the Python server and updates the UI dynamically based on server responses.
    \item \textit{Core Functionalities}: Processes frames captured from the webcam (e.g., 10 seconds of video at 0.1 ms intervals, generating 1000 frames) and sends frames to the Python backend for emotion analysis.
    \item \textit{Dynamic Rendering}: Updates the UI in real-time based on detected emotions. For example, if the user’s emotion is detected as "angry," the system dynamically modifies the front-end layout or visuals using rendering commands.
\end{itemize}

\begin{figure*}[t] % Use [t] to place the figure at the top
    \centering
    \includegraphics[width=\textwidth]{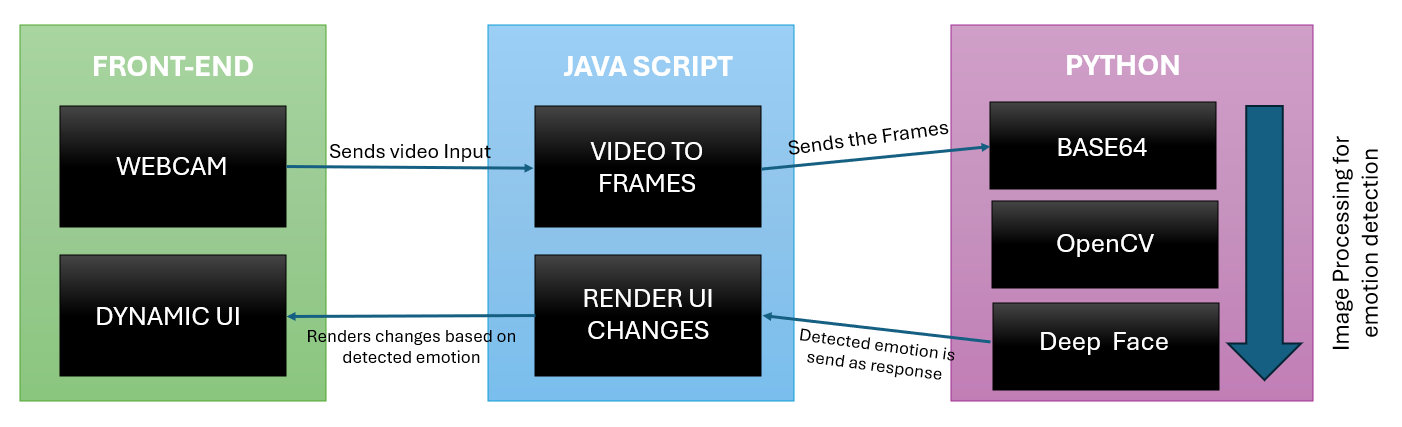} % Ensures the image spans the full width
    \caption{System Architecture for Emotion Detection using Webcam, OpenCV, and DeepFace.}
    \label{fig:system_architecture}
\end{figure*}

\subsubsection{Python Server and Emotion Detection}
The backend is powered by Python for its flexibility and comprehensive libraries for AI and emotion detection. The following key components are used:
\begin{itemize}
    \item \textit{Base64 Encoding}:
    \begin{itemize}
        \item Converts binary image data (e.g., JPEG or PNG) into text-based formats for efficient transmission between the client and server.
        \item Ensures compatibility across web systems and enables seamless decoding for further processing.
    \end{itemize}
    \item \textit{OpenCV}\cite{ED-DWT}:
    \begin{itemize}
        \item Performs preprocessing on video frames, including grayscale conversion, noise cancellation, and face recognition using Haar cascades or DNN modules to extract the region of interest (ROI).
        \item Prepares cleaned and cropped facial data for emotion detection.
    \end{itemize}
    \item \textit{DeepFace}\cite{ED-transferlearning-DL}:
    \begin{itemize}
        \item Utilizes pre-trained deep learning architectures like VGG-Face, Google FaceNet, OpenFace, and DeepID.
        \item Workflow for Emotion Detection:
        \begin{itemize}
            \item Extracted faces are aligned (e.g., ensuring the eyes are level) and resized to match input dimensions required by the deep learning models.
            \item The adjusted faces are passed through emotion recognition models, which output probabilities for emotion categories (e.g., happy, sad, angry, neutral).
            \item The emotion with the highest probability is selected as the user's detected emotion.
        \end{itemize}
    \end{itemize}
\end{itemize}
These libraries were selected after thoroughly reviewing their functionality, performance, and precision, as well as analyzing various research findings and results\cite{ED-deepface,ED-opencv}.
\subsubsection{Emotion-Based Adaptive Features}
\begin{figure}[H]
\centering
\includegraphics[width=7cm]{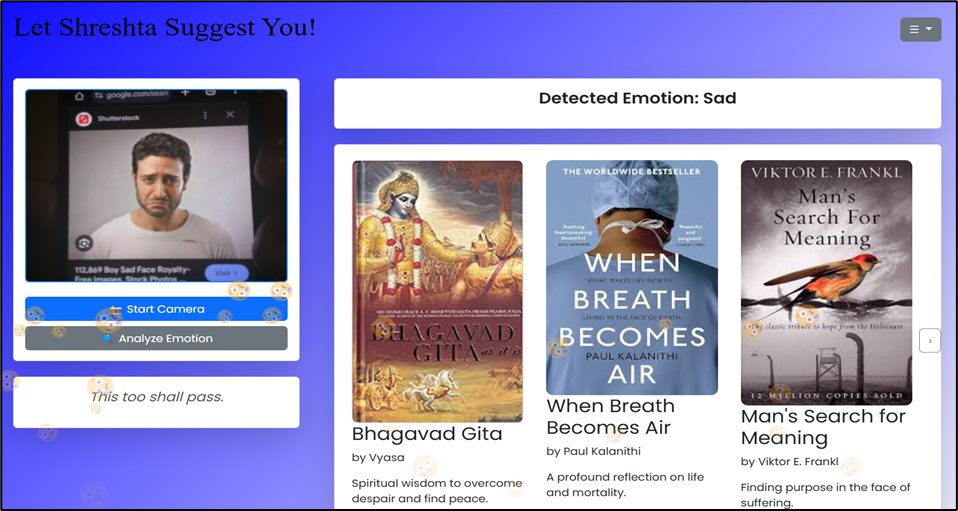}
\caption{UI changes when user is Sad}
\label{fig2}
\end{figure}

\begin{figure}[H]
\centering
\includegraphics[width=7cm]{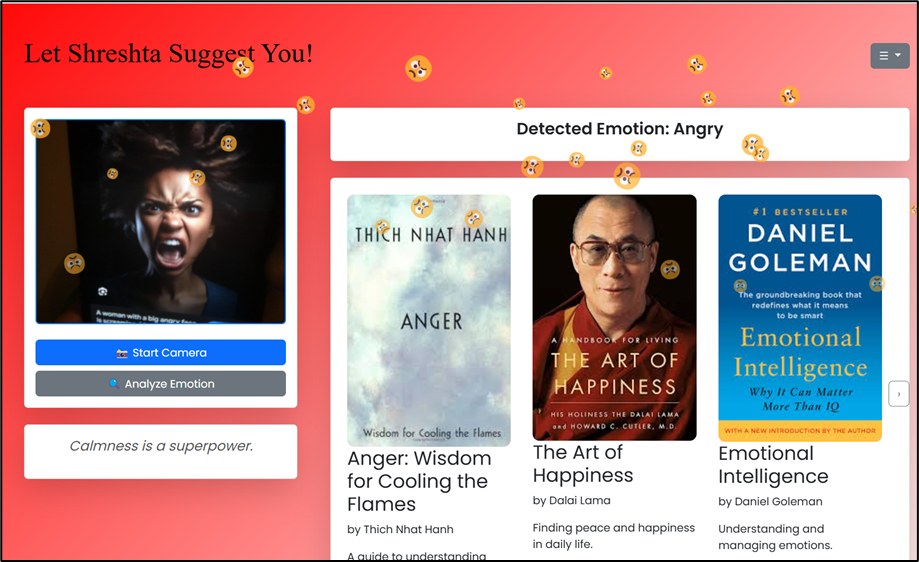}
\caption{UI chenages when the user is Angry}
\label{fig3}
\end{figure}

The system can accomplish the following key features based on the detected emotions:
\begin{itemize}
    \item The system dynamically updates UI components, including background themes, animations, and book recommendations.
    \item Emotion-based quotes are displayed to enhance user engagement, as shown in Figure \ref{fig2} and Figure \ref{fig3}.
\end{itemize}

\begin{table}[h]
\centering
\caption{The default changes on the UI based on predicted emotion}
\label{tab:ui-emotion-changes}
\begin{tabular}{|c|p{7cm}|}
\hline
\textbf{Emotion} & \textbf{Changes in UI (default)} \\ \hline
Happy & 
\begin{itemize}
    \item Background Yellow
    \item Happy emojis raining down
    \item Feel good books for happy readers are suggested
    \item A quote that resembles happiness is displayed
\end{itemize} \\ \hline
Sad & 
\begin{itemize}
    \item Pale blue background color
    \item Sad emojis raining down
    \item Inspirational and motivational books suggested for readers facing depression
    \item A motivational quote
\end{itemize} \\ \hline
Angry & 
\begin{itemize}
    \item Red background
    \item Angry emojis falling like rain
    \item Books related to anger management
    \item Anger management quote
\end{itemize} \\ \hline
Neutral & 
\begin{itemize}
    \item Gray background
    \item Normal emojis falling like rain
    \item Feel good books, with neutral emotion like autobiographies
    \item Message saying ``balance is key''
\end{itemize} \\ \hline
Surprised & 
\begin{itemize}
    \item Pink background
    \item Shocking emojis falling like rain
    \item Thrillers, fantasy, sci-fi books
    \item Message saying ``I love surprises''
\end{itemize} \\ \hline
\end{tabular}
\end{table}

\begin{figure*}[t] % Use [t] to place the figure at the top
    \centering
    \includegraphics[width=\textwidth]{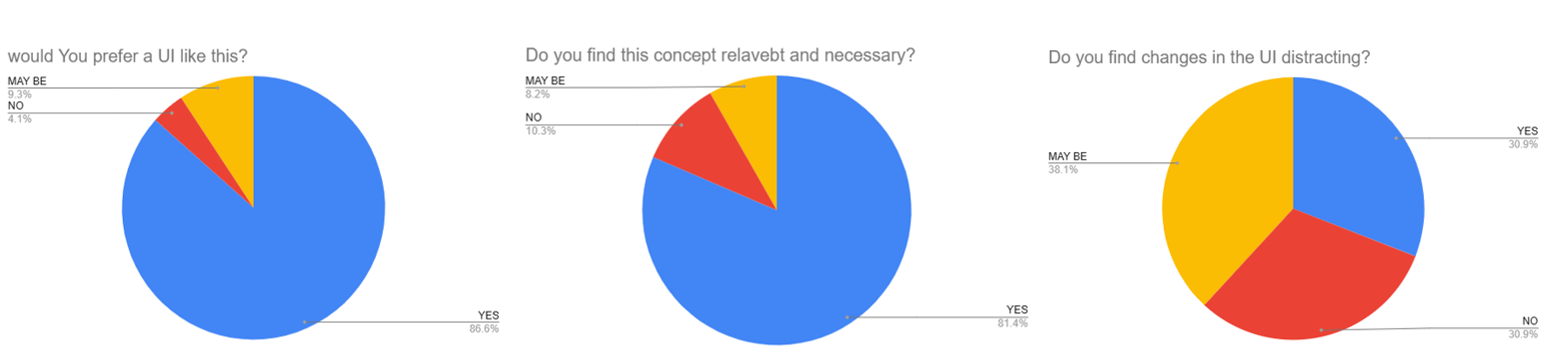} % Ensures the image spans the full width
    \caption{Results from the Survey conducted}
    \label{fig:system_architecture}
\end{figure*}

\subsection{Additional Features of this application}
\begin{figure}[H]
\centering
\includegraphics[width=7cm]{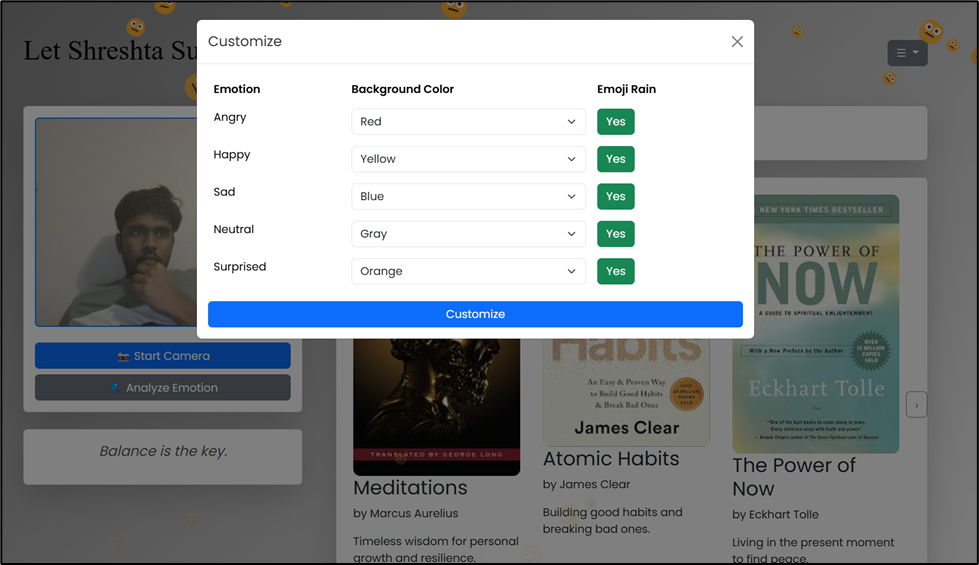}
\caption{UI Customization menu}
\label{fig4}
\end{figure}

The implemented Dynamic UI is set to default changes when the user begins interacting with the system; however, the default configuration may not suit all users. For instance, some users might prefer a green background color when sad, accompanied by soft background music and no animations in the UI. To accommodate such preferences, the application provides custom UI options, allowing users to modify settings according to their emotional state. Additionally, users can disable animations if they find them unappealing. Figure \ref{fig4} shows us the UI customization menu where the user can make changes as per their requirements\cite{customUI}.

\begin{figure}[H]
\centering
\includegraphics[width=7cm]{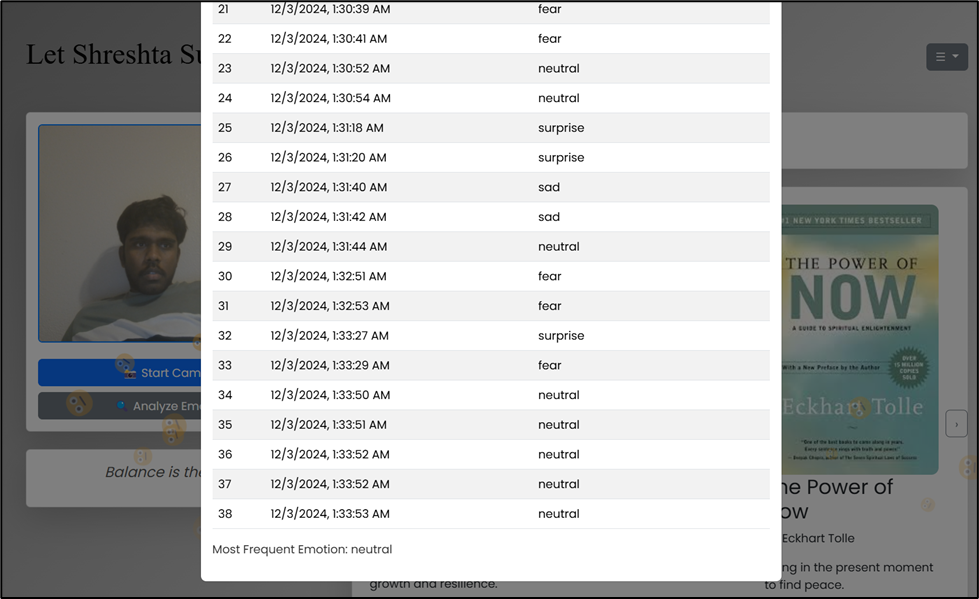}
\caption{Emotion tracking}
\label{fig5}
\end{figure}

The application also tracks user emotions during usage. As a user engages with the system regularly over a month, the UI not only adapts changes to the front-end but also monitors and records the user's emotional patterns. It calculates the most frequent emotions, providing insights into user behavior, which is shown in Figure \ref{fig5}. This method combines emotional data with quantitative usage metrics to swiftly and effectively evaluate user acceptance. Emotional tracking supports agile development by enabling continuous evaluation of app usability and user satisfaction alongside ongoing development\cite{emotion-tracking}.

Currently, the application is limited to manual customization and basic emotion tracking. Future enhancements will integrate advanced AI and emotion tracking capabilities, enabling the UI to adapt dynamically without requiring user intervention. This integration of AI will facilitate autonomous adjustments, creating a seamless and personalized user experience without reliance on a customization menu.

\subsection{Survey Results}
The survey was conducted using Google Forms, where participants were asked a series of questions to understand their perspectives on such applications. The questions included:

\begin{itemize}
    \item Would you prefer a UI like this? (Yes/No/Maybe)
    \item Do you find the changes in the UI distracting? (Yes/No/Maybe)
    \item According to current technological requirements, do you find this concept relevant and necessary? (Yes/No/Maybe)
    \item Overall, how would you rate such an application on a scale of 1 to 5?
\end{itemize}

The survey received 97 responses from individuals across various age groups and professions, including students, teachers, working professionals, and individuals from non-technical backgrounds. This diversity ensures that the survey is not focused on a particular profession or section of society. Additionally, the gender ratio of respondents was approximately 52:48 (men to women), ensuring minimal gender bias in the results.

The survey results clearly indicate a high demand for such applications, with approximately 87\% of respondents preferring a dynamic, interactive UI over a static one. These respondents appreciated the UI's ability to adapt to their needs and emotions. However, a small number of participants, accounting for less than 5\% of the total responses, expressed a preference against such a UI. Further analysis revealed that this group primarily consisted of working professionals who found an interactive UI potentially distracting in their work environments.

The question, "Do you find the changes in the UI distracting?" received mixed responses, with a significant proportion of users selecting a neutral option. Although the majority of respondents did not find the changes distracting, a notable number felt unsure or found the changes distracting. To address these concerns, the application includes a customization menu that allows users to tailor the UI to their preferences. Additionally, future integration of AI will further reduce distractions and enhance user engagement by making the UI more adaptive and context-aware. This approach ensures that the application remains suitable for a broad range of users while minimizing potential drawbacks associated with interactive UI designs.
This research involved conducting a survey with participants recruited from a global pool of college students and working professionals. These subjects were engaged through online platforms and asked to complete a Google Forms survey relevant to our study on emotion-based adaptive user interfaces.
\subsection{Ethical Impact}
IRB and Ethical Approval: Given the nature of the survey, which was non-invasive and collected data without any personal identifiers, it did not require Institutional Review Board (IRB) approval. The participants were not affiliated with our institution, which further limits our ability to seek IRB oversight. However, ethical considerations were thoroughly evaluated to ensure no harm to the participants and to maintain the integrity of the research process.

Informed Consent: All participants were provided with clear information about the purpose of the survey, its voluntary nature, and the anonymity of their responses prior to participation. They consented to participate under these conditions, ensuring transparency and respect for their autonomy.

Compensation: No compensation was offered or provided to the participants, as their involvement required minimal effort and time, and there was no risk associated with their participation.

Our findings are based on a limited sample that may not fully represent the global population. This limitation affects the scalability and applicability of our results across different demographic groups.

The performance of our emotion recognition technology may vary across different contexts, particularly affecting non-native English speakers or cultural expressions of emotion that differ from those represented in the training data.

\section{conclusion}
Emotion-aware adaptive UIs, such as Face2Feel, are crucial to addressing the growing demand for technological advancements. An adaptive UI enhances user interaction with the system, making it more tailored to their needs. This paper presents Face2Feel, a practical implementation of an emotion-aware adaptive UI system, demonstrating its feasibility and potential to improve user experience. 

Research indicates a growing correlation between increased electronic gadget usage, social isolation, and mental health challenges \cite{suicide,Adolesent-health}. Emotion-aware adaptive UIs, such as Face2Feel, offer a promising approach to mitigate these concerns by enhancing user interaction and personalization. By making social media and other platforms of communication more engaging and user-friendly, adaptive UIs can significantly improve user experience. Beyond these points, emotion-adaptive UIs have broader applicability, such as chatbots that navigate complex websites, assist users in finding desired resources, or provide answers to FAQs in the banking sector and corporate environments. Similarly, emotion-aware systems can be utilized for customer feedback and review systems, recommendation engines, and other applications that require a personalized touch.

This study demonstrates the potential of emotion-aware adaptive UIs to create more interactive, efficient, and user-centric digital environments, paving the way for their broader implementation across various industries.
\bibliographystyle{IEEEtran} % Or any other preferred style
\bibliography{references}

\begin{thebibliography}{10}
\providecommand{\url}[1]{#1}
\csname url@samestyle\endcsname
\providecommand{\newblock}{\relax}
\providecommand{\bibinfo}[2]{#2}
\providecommand{\BIBentrySTDinterwordspacing}{\spaceskip=0pt\relax}
\providecommand{\BIBentryALTinterwordstretchfactor}{4}
\providecommand{\BIBentryALTinterwordspacing}{\spaceskip=\fontdimen2\font plus
\BIBentryALTinterwordstretchfactor\fontdimen3\font minus \fontdimen4\font\relax}
\providecommand{\BIBforeignlanguage}[2]{{%
\expandafter\ifx\csname l@#1\endcsname\relax
\typeout{** WARNING: IEEEtran.bst: No hyphenation pattern has been}%
\typeout{** loaded for the language `#1'. Using the pattern for}%
\typeout{** the default language instead.}%
\else
\language=\csname l@#1\endcsname
\fi
#2}}
\providecommand{\BIBdecl}{\relax}
\BIBdecl

\bibitem{suicide}
M.-T. C, W.~M, S.~M, C.~J-D, B.~S, and L.~C, ``Social isolation and suicide risk: Literature review and perspectives,'' \emph{European Psychiatry}, vol. e65, 2022.

\bibitem{Adolesent-health}
McLeod, J.~D., U.~R, and R.~S, ``Adolescent mental health, behavior problems, and academic achievement,'' \emph{Journal of Health and Social Behavior}, vol. 53(4), 2012.

\bibitem{EBAUI}
M.~Alipour, M.~Tourchi~Moghaddam, K.~Vaidhyanathan, and M.~Baun~Kj\ae{}rgaard, ``Toward changing users behavior with emotion-based adaptive systems,'' \emph{Association for Computing Machinery}, vol.~7, 2023.

\bibitem{intelligent_UI_design}
T.~Song, X.~Li, B.~Wang, and L.~Han, ``Research on intelligent applicationdesignbased on artificial intelligence andadaptiveinterface,'' \emph{World Journal of Innovation and Modern Technology}, vol.~7, 2024.

\bibitem{Emoticontrol}
M.~Alipour, M.~T. Moghaddam, K.~Vaidhyanathan, , and M.~B. Kjærgaard, ``Emoticontrol: Emotions-based control of user-interfaces adaptations,'' \emph{Association for Computing Machinery}, vol.~7, 2023.

\bibitem{Model-based-UI}
J.~Hussain1, A.~U. Hassan1, H.~S.~M. Bilal1, R.~Ali2, M.~Afzal3, S.~Hussain1, J.~Bang1, O.~Banos4, and S.~Lee1, ``Model-based adaptive user interface based on context and user experience evaluation,'' \emph{J Multimodal User Interfaces}, vol.~12, 2018.

\bibitem{component-based-AUI}
\BIBentryALTinterwordspacing
E.~Yigitbas, K.~Josifovska, I.~Jovanovikj, F.~Kalinci, A.~Anjorin, and G.~Engels, ``Component-based development of adaptive user interfaces,'' in \emph{Proceedings of the ACM SIGCHI Symposium on Engineering Interactive Computing Systems}, ser. EICS '19.\hskip 1em plus 0.5em minus 0.4em\relax New York, NY, USA: Association for Computing Machinery, 2019. [Online]. Available: \url{https://doi.org/10.1145/3319499.3328229}
\BIBentrySTDinterwordspacing

\bibitem{CBAUI-best-approach}
\BIBentryALTinterwordspacing
U.~Hub, ``Enhancing ui development with a component-based approach,'' 2023, accessed: 2024-12-31. [Online]. Available: \url{https://uihub.licode.ai/blog/enhancing-ui-development-with-component-driven-user-interfaces-approach}
\BIBentrySTDinterwordspacing

\bibitem{webcomponents}
T.~Bui, \emph{Web Components: Concept and Implementation}.\hskip 1em plus 0.5em minus 0.4em\relax Self-Published, 2019.

\bibitem{dynamic-html}
D.~Goodman, \emph{Dynamic HTML: The Definitive Reference: A Comprehensive Resource for HTML, CSS, DOM \& JavaScript}.\hskip 1em plus 0.5em minus 0.4em\relax O'Reilly Media, Inc., 2002.

\bibitem{Mahdi}
M.~H. Miraz, M.~Ali, and P.~S. Excell, ``Adaptive user interfaces and universal usability through plasticity of user interface design,'' \emph{Computer Science Review}, vol.~40, 2021.

\bibitem{emotion-tracking}
P.~Mennig, S.~A. Scherr, and F.~Elberzhager, ``Supporting rapid product changes through emotional tracking,'' \emph{IEEE/ACM 4th International Workshop}, 2019.

\bibitem{ED-DWT}
R.~I. Bendjillali, M.~Beladgham, K.~Merit, and A.~Taleb-Ahmed, ``Improved facial expression recognition based on dwt feature for deep cnn,'' \emph{Electronics}, vol.~8, 2019.

\bibitem{ED-opencv}
N.~Boyko, O.~Basystiuk, and N.~Shakhovska, ``Performance evaluation and comparison of software for face recognition, based on dlib and opencv library,'' \emph{IEEE Second International Conference on Data Stream Mining \& Processing (DSMP)}, 2018.

\bibitem{ED-transferlearning-DL}
M.~A.~H. Akhand, S.~Roy, N.~Siddique, M.~A.~S. Kamal, and T.~Shimamura, ``Facial emotion recognition using transfer learning in the deep cnn,'' \emph{Electronics}, vol.~10, 2021.

\bibitem{ED-deepface}
A.~Awana, S.~V. Singh, A.~Mishra, V.~Bhutani, S.~R. Kumar, and P.~Shrivastava, ``Live emotion detection using deepface,,'' \emph{6th International Conference on Contemporary Computing and Informatics (IC3I)}, 2023.

\bibitem{multiprocessing-timecomplexity}
D.~Sarkar, ``Cost and time-cost effectiveness of multiprocessing,'' \emph{IEEE Transactions on Parallel and Distributed Systems}, vol.~4, no.~6, pp. 704--712, 1993.

\bibitem{multiprogramming}
M.~Crovella, P.~Das, C.~Dubnicki, T.~LeBlanc, and E.~Markatos, ``Multiprogramming on multiprocessors,'' \emph{Proceedings of the IEEE International Conference on Distributed Computing Systems}, pp. 590--597, 1991.

\bibitem{bootstrap}
Spurlock and Jake, \emph{Bootstrap: responsive web development}.\hskip 1em plus 0.5em minus 0.4em\relax O'Reilly Media, Inc., 2013.

\bibitem{customUI}
S.~L.~T. Hui and S.~L. See, ``Enhancing user experience through customisation of ui design,'' \emph{Procedia Manufacturing}, vol.~3, 2015.

\end{thebibliography}
\end{document}